\documentstyle[11pt,epsfig,twocolumn]{article}
\title{{\bf Remarks on perturbation theory for Hamiltonian systems
}}
\author{{\bf Alexander Rauh}\\
{\small  Fachbereich Physik, Carl von Ossietzky Universit\"at,
D-26111 Oldenburg, Germany }}

\date{}

\begin{document}
\maketitle

\newcommand{\beq}{\begin{equation}}                                            
\newcommand{\beqa}{\begin{eqnarray}}                                         
\newcommand{\eeq}{\end{equation}}          
\newcommand{\eeqa}{\end{eqnarray}}          
\newcommand{\pa}{\partial}                                                    
\newcommand{\jm}{\jmath}

\begin{abstract}
A comparative discussion of the normal form and action angle variable
method is presented in a tutorial way. Normal forms are introduced
by Lie series which avoid mixed variable canonical transformations.
The main interest is focused on establishing a third integral of
motion for the transformed Hamiltonian truncated at finite order
of the perturbation parameter. In particular, for the case of the
action angle variable scheme, the proper canonical transformations
are worked out which reveal the third integral in consistency with
the normal form. Details are discussed exemplarily for the 
H\'{e}non-Heiles Hamiltonian.  The main conclusions  are generalized 
to the case  of $n$ perturbed harmonic oscillators.  
\end{abstract}

\section{Introduction}

The following contribution is concerned with finite perturbation series
characterized by  bounded remainders in properly chosen compact domains 
of phase space. We are, however, not interested here in  estimating the 
rest terms.  As has been  known since Poincar\'{e}, infinite perturbation 
series  for Hamiltonians, in general, do not converge in compact domains.  
Or more precisely, if convergence takes place according to the KAM theorem, 
then it generally occurs in an invariant subset of  phase space whose
complement is open and dense \cite{antonio}.
 
Let us assume that the Hamiltonian can be brought into the  form 
$H=h+\epsilon V$ where $h$ refers to $n$ uncoupled harmonic 
oscillators and $\epsilon$ is the smallness parameter. Then the 
main differences of a perturbative treatment by normal forms
and mixed variable generating functions, respectively,  can be 
characterized as follows. In the latter method  one uses action angle 
variables $I\in {\bf R}^n,\phi\in {\bf T}^n$ and tries to find a canonical 
transformation $(I,\phi)\rightarrow (J,\psi)$ which makes the
transformed Hamiltonian independent of $\psi$ in a certain domain $D$ of
phase space. For an elementary introduction into this method,
including the main ideas of the proof of the KAM theorem, the textbook 
\cite{rasetti} is recommended.
 In the normal form case, on the other hand, one adopts
complex canonical variables $(u_{\nu},v_{\nu})$,
and tries to make $H$ canonically equivalent to $n$ harmonic oscillators
given by
\beq
h=-i \sum_{\nu=1}^n  \, \omega_{\nu} u_{\nu}v_{\nu}.
\eeq
 Both strategies fail, because
terms with  resonance denominators occur to any order, in general,
 which cannot be transformed away.
Thus, even if one accepts a finite cutoff at order $\epsilon^N$, it is not 
possible, in general, to transform a Hamiltonian into an integrable form.
There seems to be at least one advantage with normal forms: they 
straightforwardly provide us
with a third integral.  As a  consequence,  in the case
of two degrees of freedom, for instance, the cutoff part of the Hamiltonian
which is normalized up to order $N$, is integrable within the definition
domain of the normal form transformation. In the case of the 
H\'{e}non-Heiles Hamiltonian  \cite{henon} this was first demonstrated  by 
\cite{gustavson}.
 As a further advantage, the 
normal form transformation can be carried out very efficiently by 
Lie series and thus by symbolic computer algebra \cite{antonio}. 

As a power series in $\epsilon$, both perturbation schemes should
 be equivalent. However, it does not  seem to be obvious,
how the third integral  can be detected
in the action angle variable picture. We will adopt an iterated 
transformation scheme where   new canonical variables are introduced at 
each perturbation order. Eventually, an elementary linear canonical
transformation will make  coming  forth the additional integral. 
For demonstration,  the H\'{e}non-Heiles 
 Hamiltonian is considered. The results are generalized 
to the case of $n$ perturbed harmonic oscillators.

 The existence of a third
integral up to order $\epsilon^N$ may be useful in Nekhoroshev-like
estimates \cite{nekhoroshev},\cite{nekhoroshev2}, \cite{galgani}.
 For instance, in the
three body problem  an approximate integral, in addition to energy
and angular momentum, should help to get sharper bounds for the 
remainders. We make use of this occasion to remark, that a former study by
the present  author \cite{rauh} on
the three body problem in celestial mechanics, essentially, was a failure
because of explicitly and tacitly (eq.(77)) adopted 
adiabatic assumptions;  the rigorous estimates in the Appendices C and D of
\cite{rauh}, on the other hand, may be helpful elsewhere for similar problems.
 The N-body problem of celestial mechanics was recently  
examined more rigorously by
\cite{niederman} within the Nekhoroshev scheme.

In the next section we will briefly introduce to the Lie series formulation
of time dependent perturbation theory for Hamiltonian systems. 
Most numerical integrators are based on time 
series with small time step. Section 2 also serves to prepare
the Lie series method for normal forms in section 3. In section 4 the 
perturbation method by action angle variables will be discussed.
In appendix A we present a proof of a  recursive formalism
for Lie series \cite{antonio},   and  in appendix B  an exact
canonical transformation is presented which brings the  
H\'{e}non-Heiles  Hamiltonian into an integrable form up to the
second order of the perturbation parameter.

\section{Time dependent perturbation}

We assume finite dimensional systems.
The Hamiltonian field generates a flow which, during the time $t$, maps a 
phase space domain $D_0$ of initial points into the domain $D_t$. If the 
field is sufficiently smooth  in the image domains  
$D_{\tau}$ for $\tau \in (0,t)$, then $D_t$ is diffeomorph to
$D_0$. In particular, inner points of $D_0$ are mapped into inner
points of $D_t$, and the boundaries of $D_t$ and $D_0$ are equally  smooth.
Moreover, by the Liouville theorem, the domain volumes are 
preserved. Chaotic behaviour, clearly, develops for such systems, if at all,
as an asymptotic property. Its observability depends
 on the adopted degree of resolution.

Lie series are tied to the time evolution of a function $f(p,q)$ where
$\{p,q\}\equiv \{p(t;p_0,q_0),$
$ p(t;q_0,q_0)\}$ are the generalized 
momentum and position at time $t$ of a trajectory with initial point 
$\{p_0,q_0\} \in {\bf R}^{2 n}$. We are interested in the development
of $f$ along a trajectory with given initial point, and write therefore
$f(p,q)=:F(t;p_0,q_0)$ with $F(0;p_0,q_0)=f(p_0,q_0)$. The Taylor expansion
in the time interval $t\in (0,\Delta t)$  reads
\beq
F(t)=\sum^N_{k=0}\frac{t^k}{k!}\, F^{(k)}(t=0) +O([\Delta t]^{N+1}).
\eeq
 Defining the linear operator $L_H$ by the Poisson bracket
\beq
L_H\,f :=\sum^n_{k=1} \left (\frac{\pa H}{\pa p_k}\,\frac{\pa f}{\pa q_k}
-\frac{\pa H}{\pa q_k}\,\frac{\pa f}{\pa p_k}\right)
\eeq
and making use   of the canonical equations, we can write
\begin{eqnarray}
\frac{df}{dt}=\sum^n_{k=1} \left (\frac{\pa f}{\pa p_k}\dot{p}_k+
\frac{\pa f}{\pa q_k}\dot{q}_k\right )
\nonumber\\
=
\sum^n_{k=1} \left (\frac{\pa f}{\pa p_k}(-\frac{\pa H}{\pa q_k})+
\frac{\pa f}{\pa q_k}\frac{\pa H}{\pa p_k}\right )=L_H\,f.
\end{eqnarray}
As a consequence, the Taylor expansion can be expressed in terms of 
the following Lie series
\beq 
F(t)=\sum^N_{k=0}\frac{t^k}{k!}\, \left (L^k_H\, f(p,q)\right )
_{(p,q)=(p_0,q_0)} +O([\Delta t]^{N+1}),
\eeq
and in the limit $N\rightarrow \infty$ we can write in compact form
\beq
F(t)=\exp[t\,L_H]\,f(p,q)_{(p,q)=(p_0,q_0)}.
\eeq 
For small enough time steps $\Delta t$, the time evolution of any
dynamical variable, in particular $f\equiv p$ or $f\equiv q$, can be 
approximated by iterated truncated Lie series.
A cutoff at $N=4$ corresponds to a fourth order Runge-Kutta
integration. If $H$ and $f$ are given analytically, the coefficients
of the Lie series can also be determined analytically, e.g. by
means of symbolic computer calculators. However, in each step symplecticity
is fulfilled only up to an error of order $[\Delta t]^{N+1}$.
In numerical problems, it may be more adequate to adopt so-called
symplectic integrators which are canonical in every step within the
number precision of the computer, see e.g. \cite{yoshida}. 
The remainder of the truncated Lie series, on the other hand, 
can be rigorously expressed through (if $f$ is  scalar) 
\beq
O([\Delta t]^{N+1})=\frac{F^{N+1}(t^*)}{(N+1)!}\, [\Delta t]^{N+1}
\hspace{0.15cm} \rm{with}\hspace{0.15cm} t^*\in (0,\Delta t),
\eeq
or
\beq
O([\Delta t]^{N+1})=\frac{f^{N+1}(p^*,q^*)}{(N+1)!}\, [\Delta t]^{N+1}
\eeq
with $(p^*,q^*)=(p(t^*),q(t^*)).$
From a priori or a posteriori knowledge on the domain of $(p,q)$,
the remainder can be estimated by upper bounds.

\section{Normal form and third integral}

For demonstration, let us consider the H\'{e}non-Heiles Hamiltonian 
\cite{henon}
\begin{eqnarray}
H=h+\epsilon\, V \hspace{0.5cm} \rm{with} \hspace{0.5cm}
h&=&\frac{1}{2} (p_1^2+q_1^2+p_2^2+q_2^2), \nonumber\\
V&=&q_1 q_2^2-\frac{1}{3} q_2^3
\end{eqnarray}
where $(p_{\nu},q_{\nu})\in {\bf R}^2$ with  $\nu=1,2$ are canonical variables.
We assume that the variables have been made dimensionless, in particular
H=1, which implies that the smallness parameter $\epsilon$ is proportional
to the square root $\sqrt{E}$ of the energy  of a given trajectory.
With the aid of the linear canonical transformation $(p_{\nu},q_{\nu})$
 $\rightarrow (u_{\nu},v_{\nu})$ where
\beq
u_{\nu}=(q_{\nu}-i\, p_{\nu})/\sqrt{2},  \hspace{.3cm}
v_{\nu}=i\,(q_{\nu}+i\, p_{\nu})/\sqrt{2},
\eeq
we obtain
\begin{eqnarray}
h&=&-i\,(u_1 v_1+u_2 v_2); 
\\
V&=&\frac{u_2-i\,v_2}{2\sqrt{2}}\left [ (u_1-i\,v_1)^2-
\frac{1}{3}(u_2-i\,v_2)^2 \right ],
\nonumber\end{eqnarray}
which has the suitable form for being subject to a normal form 
transformation.
 
In section 2, from the time evolution $(p_0,q_0)$
 $\rightarrow (p(t),q(t))$, we had derived
 as generator of a canonical transformation the operator $\exp[t\,L_H]$.
 Clearly, any function $H(p,q)$ which does not explicitly depend on time, 
 gives rise to such a generator. Moreover, the variable $t$ in  
$\exp[t\,L_H]$ does not need to be identified as time, it can be any real 
parameter. This is seen e.g., when symplecticity is inferred from the 
Poisson bracket  $L_p\,q$ calculated with respect to  
$(p_0,q_0)$. It is  therefore legitimate to adopt $\epsilon$ as parameter 
of the generating function \cite{deprit}.
The point of view adopted here is to generate, at any given time $t$, a 
canonical transformation $(p(t;\epsilon=0),q(t;\epsilon=0))$ $\rightarrow$ 
$(p(t;\epsilon),q(t;\epsilon))$  , 
which is parametrized with respect to
the interaction parameter $\epsilon$. The transformed Hamiltonian 
is written as a power series 
\beqa
H(p(t;\epsilon),q(t,\epsilon))=h(p(t;0),q(t;0))+
\nonumber \\
\sum^{\infty}_{k=1}H_k(p(t;0),q(t;0))
 \epsilon^k
\eeqa
with $h$ being the unperturbed Hamiltonian. Clearly, such
canonical transformations can be achieved by means of 
arbitrary scalar functions. In the following it is convenient to write
the generating function in the form
\begin{eqnarray}
\chi'(p(t;\epsilon),q(t;\epsilon)):
&=&\frac{d}{d\epsilon} \chi(p(t;\epsilon),q(t;\epsilon))
\nonumber\\
&=&
\frac{\pa \chi}{\pa p}\frac{d p}{d\epsilon}+
\frac{\pa \chi}{\pa q}\frac{d q}{d\epsilon}
\eeqa
where we assume
\beq
\frac{d p}{d\epsilon}=-\frac{d \tilde{h}}{d q};\hspace{0.5cm}
\frac{d q}{d\epsilon}=\frac{d \tilde{h}}{d p}
\eeq
for some scalar function $\tilde{h}(p,q)$ which we do not need to specify.

The  canonical transformation of  an arbitrary scalar function $g$ 
is defined by the 
constituent equation (we omit writing the time parameter)
\beq
\frac{d}{d\epsilon} g(p(\epsilon),q(\epsilon))=
L_{\chi'}\, g(p(\epsilon),q(\epsilon)),
\eeq
which gives rise to the power series representation
\beq
g(p(\epsilon),q(\epsilon))=\exp[\epsilon\,L_{\chi_0'}]\, 
g(p(0),q(0))
\eeq
with $\chi_0'=\chi'(p(0),q(0))$.

When $\chi$, too, is expanded in a power series
\beq
\chi(p(\epsilon),q(\epsilon))=\sum^{\infty}_{k=1}\epsilon^k
\chi_k(p(0),q(0)),
\eeq
then the arbitrary functions $\chi_k$ will be at our disposition to simplify
 the Hamiltonian coefficients $H_k$. Henceforth we will write simply
$(p,q)$ for the phase space variables
$(p(\epsilon=0),q(\epsilon=0))$. 
As is shown in Appendix A, the transformed terms, $H_k$, can be determined
recursively as follows \cite{antonio}
\begin{eqnarray}
H_0 & = & h; \hspace{.3cm} L_{h}\,\chi_1+H_1=V_0; \\
L_h\, \chi_k+H_k &= &\frac{1}{k}\,
V_{k-1}+\sum_{j=1}^{k-1}\frac{j}{k}L_{\chi_j}\,H_{k-j}, 
\end{eqnarray}
$k=2,3,...$, where $V_j$ is defined through the power series of the
transformed  potential, namely  \linebreak $\exp[\epsilon\,L_{\chi'}]\, V = 
\sum_{j=0,1,..}V_j \, \epsilon^j $.

Let us start with the term $k=1$. Then we have, with $V_0\equiv V$,
\beq
L_h\, \chi_1+H_1=V,
\eeq
and we try to set $H_1=0$ with the implication that $\chi_1$ has to fulfil
the relation $L_h\, \chi_1=V$. To discuss, whether $V$ is in the range
of the homology \cite{arnold}
 operator $L_h$, we adopt the canonical variables (10)
together with the representation  (1) of $h$. Furthermore, we  
exploit the fact that $V$ is a linear combination of monomials of the form
$U^m:=u_1^{m_1}v_1^{m_2}u_2^{m_3}v_2^{m_4}$ with $|m|:=m_1+m_2+m_3+m_4$
$= 3$ and $m_j \in {\bf N}_0$ for $j=1,2,3,4$.
Now, each monomial is an eigenfunction of $L_h$, because
\beqa
L_h\, U^m&\equiv& \sum_{\nu=1,2}\left (\frac{\pa h}{\pa u_{\nu}}
\frac{\pa U^m}{\pa v_{\nu}}-
\frac{\pa h}{\pa v_{\nu}}\frac{\pa U^m}{\pa u_{\nu}} \right )
\nonumber\\&=&
i(m_1-m_2+m_3-m_4)U^m.
\eeqa
As a consequence, the set of resonance monomials defined by
\beq
\{U^m\,\,|\,\,m_1+m_3 =  m_2+m_4=0;\,\,\, m_i\in {\bf N}_0\,\}
\eeq
 are not in the range of $L_h$,
and therefore cannot be removed by the generating function $\chi_1$.
Clearly, the resonance case is possible for monomials of even
order $|m|$ only. Since $V$, according to (11), consists of third order terms,
eq.(20) is solvable for $\chi_1$ with $H_1$ set equal to zero. The 
general solution includes an arbitrary part of the
kernel of $L_h$ consisting of resonance
monomials. If, as usual, this kernel part is set equal to zero, $\chi_1$ is
uniquely given by  a linear combination of the monomials occurring 
in $V$.  Furthermore, $V_1=L_{\chi_1}\,V$ is now determined in terms of
4-th order monomials.  

We examine the next iteration, which will be sufficient to reveal the general
structure of the normalized Hamiltonian:
\beq
L_h\,\chi_2+H_2=\frac{1}{2}\,V_1+\frac{1}{2}L_{\chi_1}\,H_1=\frac{1}{2}\,V_1.
\eeq
Here, $V_1$ contains  both types of monomials, nonresonant ones which  
are in the range of the operator
$L_h$ and resonant monomials. The latter must be compensated by $H_2$,
while the nonresonant terms are transformed away by the proper choice of
$\chi_2$. This is typical of all orders. Thus, an optimal simplification
is achieved when 
the generating function is disposed of  in such a way that the transformed
Hamiltonian terms $H_k$ contain resonant monomials only. 

When this normalization is carried out up to order $N$, then 
 the truncated Hamiltonian
\beq
H^{(N)}:=\sum_{k=0}^N \epsilon^k\,H_k
\eeq
is a constant of motion up to a rest term of the order $\epsilon^{N+1}$.
Moreover, by the definition (22) of the resonance monomials and because
of (21), we have the property
\beq
L_h\,H^{(N)}=0
\eeq
which tells that $h$ is in involution with $H^{(N)}$
and therefore a further constant of motion. As should be remarked,
the remainder $R_{N+1}$ in general is finite within properly
chosen domains of phase space \cite{antonio}.  

Let $E$ and $h$ be the integral constants of a given trajectory. 
Going back to the original variables and choosing as Poincar\'{e}
surface of section the plane $q_2=0$, one  eliminates the variable $p_2$
from the energy integral through
\beq
p_2=p_2(p_1,q_1,q_2=0; E),
\eeq
and inserts $p_2$ into the third integral
\beq
h=h\left (p_1,q_1,p_2(p_1,q_1,E) \right ).
\eeq
The latter equation implicitly defines one-dimensional manifolds
$M(p_1,q_1;E,h)=0$ which for constant energy $E$ and different values
$h$ were first plotted in reference \cite{gustavson}. The manifolds turned out 
as closed curves
corresponding to the intersection of 2-tori with the Poincar\'{e} plane
and thus demonstrating the integrability of the approximated H\'{e}non-
Heiles Hamiltonian. For small enough energies $E\equiv \epsilon^2$,
as is well known, this picture is confirmed by numerical integration
of the model.  A compact symmetrized form of the normalized 
H\'{e}non-Heiles Hamiltonian can be found in \cite{rak}

\section{Mixed variable generating function}

In terms of  action angle variables  
$(I_{\nu},\phi_{\nu})\leftarrow (p_{\nu},q_{\nu})$ defined by
\beq
p_{\nu}=\sqrt{2\, I_{\nu}}\cos(\phi_{\nu}); \hspace{.3cm}
q_{\nu}=\sqrt{2\, I_{\nu}}\sin(\phi_{\nu}),
\eeq
$\nu=1,2$, the H\'{e}non-Heiles Hamiltonian reads
\beqa
H=h+\epsilon\, V; \hspace{0.3cm} h&=&I_1+I_2,
\nonumber \\
 V&=&V(I_1,I_2,\phi_1,\phi_2).
\eeqa
In order to reveal a third integral in the truncated part of the 
perturbatively transformed Hamiltonian, we stepwise introduce generating
functions as follows 
\beqa
F^{(n)}(J_1,J_2,\phi_1,\phi_2)=J_1\phi_1+J_2\phi_2+
\nonumber\\
\epsilon^n\, S^{(n)}(J_1,J_2,\phi_1,\phi_2),
\eeqa
$n=1,2,..,$ which implicitly define  new canonical torus variables
$(I_1,I_2,\phi_1,\phi_2)\rightarrow (J_1,J_2,\psi_1,\psi_2)$ 
through the  relations
\beq
I_{\nu}=J_{\nu}+\epsilon^n\, \frac{\pa S^{(n)}}{\pa \phi_{\nu}}; \hspace{0.5cm}
\psi_{\nu}=\phi_{\nu}+\epsilon^n\, \frac{\pa S^{(n)}}{\pa J_{\nu}}.
\eeq

In the first step, one substitutes the old action variables
in terms of  the new ones as usual to obtain
\beqa
H^{(1)}=J_1+J_2+\epsilon \left [\frac{\pa S^{(1)}}{\pa \phi_{1}}+
\frac{\pa S^{(1)}}{\pa \phi_{2}} \right ]+
\nonumber\\
\epsilon 
V(J_1+\epsilon\frac{\pa S^{(1)}}{\pa \phi_{1}},J_2+
\epsilon \frac{\pa S^{(1)}}{\pa \phi_{2}},\phi_1,\phi_2).
\eeqa
Now we try to remove the potential term $V$ to first order in $\epsilon$
by choosing 
the Fourier components of $S=\sum_{n_1,n_2\in {\bf Z}}
S_{n_1n_2}(J_1,J_2)\exp(i\,n_1\phi_1+i\,n_2\phi_2)$ as follows
\beq
S^{(1)}_{n_1n_2}(J_1,J_2)=
i \frac{V_{n_1n_2}(J_1,J_2)}{n_1\omega_1+n_2\omega_2}.
\eeq  
Here, with the unperturbed oscillator freqencies $\omega_1=\omega_2=1$,
this is possible, because resonance components of $V$ with $n_1+n_2=0$
do not exist. With this,
the  transformed Hamiltonian reads
\beq
H^{(1)}=J_1+J_2+\epsilon^2 V^{(2)}(J_1,J_2,\psi_1,\psi_2;\epsilon)
\eeq
where, due to the elimination of the old angle variables $\phi_{\nu}$ 
in terms of $\psi_{\nu}$, the potential  $V^{(2)}$ now is an 
infinite power series in  $\epsilon$. 

Proceeding to second order, with the canonical transformation  
$(J_1,J_2,\psi_1,\psi_2)$ $\rightarrow$
$(\tilde{J}_1,\tilde{J}_2,\tilde{\psi}_1,\tilde{\psi}_2)$ defined by
$F^{(2)}(\tilde{J}_1,\tilde{J}_2,\psi_1,\psi_2)$,
we obtain the transformed Hamiltonian (omitting the tilde, for simplicity) 
\beqa
H^{(2)}&=&J_1+J_2+\epsilon^2 h_2(J_1,J_2) +
\nonumber\\
&&\epsilon^2 R^{(2)}(J_1,J_2,
\psi_1-\psi_2)+
\nonumber\\
&&\epsilon^3\, V^{(3)}(J_1,J_2,\psi_1,\psi_2;\epsilon)
\eeqa
where
\beqa
\lefteqn{h_2(J_1,J_2)=\frac{1}{(2\pi)^2}}
\\
&&\int^{2\pi}_0\,d\phi_1
\int^{2\pi}_0\,d\phi_1\,\,V^{(2)}(J_1,J_2,\phi_1,\phi_2;\epsilon=0).
\nonumber\eeqa
It is important to realize that the angle dependence of 
the resonance term $R^{(2)}$ is special and given through the difference 
$\psi_1-\psi_2$,
because it contains only  Fourier  components with $n_1+n_2=0$.

As a consequence, if we truncate at second order in $\epsilon$, we can
apply 
 the linear canonical transformation $J_1,J_2,\psi_1,\psi_2$
$\rightarrow$  $J_1',J_2',\psi_1',\psi_2'$ with
\beq
J_1':=J_1+J_2; \hspace{0.2cm} J_2':=J_2;
\hspace{0.2cm} \psi_1':=\psi_1;\hspace{0.2cm} \psi_2':=\psi_2-\psi_1
\eeq
to  arrive at an effectively one-dimensional Hamiltonian
with $J_1'$ being a constant of motion.
With respect to the remaining  degree of freedom, $(J_2',\psi_2')$, 
it is standard to achieve the integrable form,
see e.g. \cite{holmes}. 
The corresponding canonical transformation is given in Appendix B. 

In order to see that this property continues to higher orders,
it will be sufficient to go one perturbative step further.
With the aid of the generating function
$F^{(3)}(\tilde{J}_1,\tilde{J}_2,\psi_1,$ $\psi_2)$ we obtain
in terms of mixed variables
\begin{eqnarray}
H^{(3)} & = & \tilde{J}_1+\tilde{J}_2+ \nonumber\\
 &\phantom{=} & \epsilon^2\left [ h_2(\tilde{J}_1,\tilde{J}_2)+
O(\epsilon^3)\right ]+ \nonumber\\
&\phantom{=} &\epsilon^2 \left[ R^{(2)}(\tilde{J}_1,
\tilde{J}_2,\psi_1-\psi_2)+O(\epsilon^3)\right ]+ \nonumber\\
&\phantom{=} & \epsilon^3 \Bigl [ \frac{\pa S^{(3)}}{\pa \psi_1}+
  \frac{\pa S^{(3)}}{\pa \psi_2}+ 
\nonumber\\
&\phantom{=} & \phantom{\epsilon^3\Bigl [}V^{(3)}(\tilde{J}_1,\tilde{J}_2,
\psi_1,\psi_2;\epsilon=0)+
\nonumber\\
&\phantom{=} &\phantom{\epsilon^3\Bigl [}O(\epsilon) \Bigr ].
\end{eqnarray}
In the last bracket the remainder of order $\epsilon$ stems from 
the expansion of $V^{(3)}(*;\epsilon)$ as a power series in $\epsilon$.
The decisive point is  that,
by the chosen $\epsilon$-dependence of the generating functions
 (30), the frequencies remain unrenormalized.
This is also the case in the normal form method.
As a consequence, we have the same resonance condition $n_1+n_2=0$.
Taking into account that $\psi_{\nu}=\tilde{\psi}_{\nu}+O(\epsilon^3)$,
we obtain the third order transformed Hamiltonian (once more omitting
the tilde) in the form  
 \begin{eqnarray}
H^{(3)}& =& J_1+J_2+\epsilon^2 h_2(J_1,J_2) +\epsilon^3 h_3(J_1,J_2)+
\nonumber \\
& \phantom{=} & \epsilon^2 R^{(2)}(J_1,J_2,\psi_1-\psi_2)+
\nonumber \\
& \phantom{=} & \epsilon^3 R^{(3)}(J_1,J_2,\psi_1-\psi_2)+ \nonumber \\
& \phantom{=} & \epsilon^4\, V^{(4)}(J_1,J_2,\psi_1,\psi_2;\epsilon)
\end{eqnarray}
with the resonance terms $R^{(2)},R^{(3)}$ depending on the angle
difference as claimed. This property, obviously, carries  to
the higher orders, and thus leads to an integrable truncated
H\'{e}non-Heiles Hamiltonian in agreement with the normal form.

In every perturbation step one has to keep track of the definition
domain of the new action variables. For instance, if the original    
variable $I_1$ is defined in the positive interval $[0, d_1]$, then
by (31) $J_1+\epsilon^n \pa S^{(n)}/\pa \phi_1$ is confined to the same
domain. As a consequence, we have to restrict $J_1$ to
$J_1\in [0,d_1-\delta^*]$ where
\beq
\delta^*=\epsilon^n \max_{\phi_1 \in 
[0,2\pi]}\frac{\pa S^{(n)}}{\pa \phi_1} \hspace{0.3cm} 
\rm{if}\hspace{0.3cm} \delta^{*}\leq d_1;
\eeq
If $\delta^* > d_1$, then the transformation is ill defined.	

\section{Generalization}

The above reasoning  can be immediately extended to the case of
$n$ perturbed harmonic oscillators. We first discuss the normal
form method.
With the generalized multi-index notation $U^m:=u_1^{m_1}v_1^{m_1'}...
u_n^{m_n}v_n^{m_n'}$, the eigenvalue relation (21) becomes
\beqa
L_h\, U^m&\equiv& \sum^n_{\nu=1}\left (\frac{\pa h}{\pa u_{\nu}}
\frac{\pa U^m}{\pa v_{\nu}}-
\frac{\pa h}{\pa v_{\nu}}\frac{\pa U^m}{\pa u_{\nu}} \right )
\nonumber\\
&=&
iU^m\, \sum^n_{\nu=1}\omega_{\nu}(m_{\nu}-m_{\nu}'),
\eeqa
which gives rise to the resonance monomials
\beq
\{U^m\,\,|\,\, \sum^n_{\nu=1}\omega_{\nu}(m_{\nu}-m_{\nu}')=0;\,\,\,
 m_{\nu}\in {\bf N}_0\,\}.
\eeq
As is remarked, even if the frequencies are all rationally independent, there
are possible resonances with $m_{\nu}=m_{\nu}'$ for $\nu$ =1,2,..n.
Since, by the normal form method, the truncated normalized Hamiltonian
$H^{(N)}$
consists of resonance monomials only, eq.(41) implies the commutation
of the Poisson bracket, namely $L_h\, H^{(N)}=0$, and thus establishes a 
third integral $h$ in addition to the energy $H^{(N)}$.

 In the action angle variable picture the Hamiltonian (1) reads
\beq
h=\omega_1 I_1+\omega_2 I_2+...\omega_n I_n.
\eeq  
It is convenient to consider the angle variables $\phi_{\nu}$ not on
the $n$-torus but on the half open $n$-cube with $\phi_{\nu}\in [0,2\pi)$.
Adopting new canonical variables $(J_1,\psi_1;...J_n,\psi_n)$ defined by 
the scaling (we assume, for simplicity, that all $\omega_{\nu}\neq 0$)
\beq
I_{\nu}=J_{\nu}/\omega_{\nu};\hspace{0.6cm}\phi_{\nu}=
\psi_{\nu}\,\omega_{\nu} \hspace{0.6cm} \nu=1,2,...n,
\eeq
we obtain $h=J_1+J_2+..J_n$, and the Fourier representations
\[
S^{(k)}=\sum_{(\mu_1,\mu_2,..\mu_n)\in {\bf Z}^n} 
S^{(k)}_{\mu_1,..\mu_n}(J_1,..J_n)\times
\]
\vspace{-.4cm}
\beq
\exp(i\,\mu_1 \omega_1 \psi_1+...
+i\,\mu_n \omega_n \psi_n); \hspace{0.2cm} \psi_{\nu}
\in [0,2\pi/\omega_{\nu}).
\eeq
From the resonance conditions $\mu_1+\mu_2+...+\mu_n=0$, we may eliminate
e.g. $\mu_1=-\mu_2-...-\mu_n$ in the phases of the resonance terms of the 
transformed Hamiltonian with the result that these terms depend on the 
following $n-1$ differences only 
\beq
\omega_2\psi_2-\omega_1\psi_1,\hspace{0.4cm}
\omega_3\psi_3-\omega_1\psi_1,\hspace{0.2cm}... \hspace{0.2cm},
\omega_n\psi_n-\omega_1\psi_1.
\eeq
Now, after the elementary canonical transformation $(J_{\nu},\psi_{\nu})$
$\rightarrow (J_{\nu}',\psi_{\nu}')$ with
\begin{eqnarray}
(J_1',\psi_1'):& = &(\sum^n_{\nu=1}\frac{\omega_1}
{\omega_{\nu}}J_{\nu},\psi_1)  \nonumber\\
(J_{\nu}',\psi_{\nu}'):& = &(J_{\nu},\psi_{\nu}-\psi_1\omega_1/\omega_{\nu}),
\end{eqnarray}
$\nu=2,3,..n,$ the resonance terms do not depend on the new angle variable
$\psi_1'$. Therefore $J_1'$ is a constant of motion of the truncated
transformed Hamiltonian in consistency with the normal form method.

As a final remark, the Hamiltonian of the three-body problem in
celestial mechanics 
(and straightforwardly also the $N$-body case) can be
expressed in terms  of suitable action angle variables which avoid
(chart dependent)
singularities at small inclinations and eccentricities, see e.g.
\cite{rauh}. It would be interesting to find out, whether a third
integral can be worked out in a finite order perturbation procedure.
This may be helpful in estimating upper bounds over finite time
intervals of the order of the age of the planetary system.

\section*{Acknowledgements}
The author thanks Frank Buss for a critical reading of the manuscript.

\section*{Appendix A:\hspace{0.5cm} Recursive Lie series}

We prove here the recursion relations (18) and (19) in a 
different way as compared with reference \cite{gg}.
First we show that the coefficients $g_k$ of an arbitrary function
$g(p(\epsilon),q(\epsilon))$ = $\sum_k g_k\,\epsilon^k$ can be 
determined by the following recursive system \cite{antonio}
\begin{eqnarray}
g_0 & = & g(p,q); \hspace{0.7cm} g_{-n}=0 \hspace{0.2cm}\rm{for} 
\hspace{0.2cm} n=1,2,..;\nonumber\\ 
g_n & = & \sum^n_{j=1}\frac{j}{n}
L_{\chi_j}g_{n-j} \hspace{.5cm}   \rm{for}  \hspace{0.2cm} n=1,2,..,
\end{eqnarray} 
where the expansion coefficients $g_k$ and $\chi_k$ have to be taken
at the point $(p,q)$ := \linebreak $(p(\epsilon=0),q(\epsilon=0))$.  
To show that (15) follows from the recursion system, we multiply 
with $\epsilon^n$ and sum over $n$
\beq
\sum^{\infty}_{n=0}\epsilon^n g_n=g_0+\sum^{\infty}_{n=1}\frac{\epsilon^n}
{n} \sum^{n}_{j=1}j\,L_{\chi_j}g_{n-j}.
\eeq
On the left hand side we have $g$. Differentiating with respect to 
$\epsilon$, transforming the double sum on the right hand side 
and making use of the fact that
$L_{\chi}$ is linear in $\chi$, we obtain
\begin{eqnarray}
\frac{d g}{d\epsilon} & = &
\sum^{\infty}_{n=1}\epsilon^{n-1}
 \sum^{n}_{j=1}j\,L_{\chi_j}g_{n-j}
\nonumber\\
&\equiv& \sum^{\infty}_{n=1}\epsilon^{n-1}
 \sum^{\infty}_{j=1}j\,L_{\chi_j}g_{n-j}
\nonumber\\
&=&
\sum^{\infty}_{m=0}
 \sum^{\infty}_{j=1}\epsilon^{m+j-1}j\,L_{\chi_j}g_{m}\nonumber \\
 & = & \sum^{\infty}_{j=1}\epsilon^{j-1}j\,
 L_{\chi_j}\sum^{\infty}_{m=0}\epsilon^{m}g_{m}
\nonumber\\
&=&
\sum^{\infty}_{j=1}\epsilon^{j-1}j\,
 L_{\chi_j}g=L_{\frac{d \chi}{d\epsilon}}\,g\equiv L_{\chi'}\,g,
\end{eqnarray}
which is (15) as was claimed.

For the final step we transform the Hamiltonian as follows
\beq
\exp[\epsilon\,L_{\chi'}]\, H =\sum_{k=0,1,..}H_k\epsilon^k,
\eeq
and on the other hand
\beqa
\exp[\epsilon\,L_{\chi'}]\,H \equiv \exp[\epsilon\,L_{\chi'}]\,
 (h+\epsilon\, V)
\nonumber\\
=\sum_{k=0,1,..}h_k \epsilon^k+\sum_{k=1,2..}V_{k-1} \epsilon^k.
\eeqa
Comparing coefficients we obtain
\beq
H_0=h_0=h; \hspace{0.5cm} H_k=h_k+V_{k-1} \hspace{0.3cm} \rm{for}
\hspace{0.3cm} k=1,2,..
\eeq
Making use of the recursion formulas (48), we can write
\beq
H_k=\sum^{k}_{j=1}\frac{j}{k}L_{\chi_j}\, h_{k-j}+
\sum^{k-1}_{j=1}\frac{j}{k-1}L_{\chi_j}\, V_{k-1-j}
\eeq
for $k=2,3,..$. Taking out the summand $j=k$ from the first sum, and combining the
remaining  sums, we find
\begin{eqnarray}
H_k = L_{\chi_k} h+\sum^{k-1}_{j=1}\frac{j}{k}L_{\chi_j}\, 
\left (h_{k-j}+\frac{k}{k-1}V_{k-1-j} \right ) \nonumber\\
 =  L_{\chi_k} h+\sum^{k-1}_{j=1}\frac{j}{k}L_{\chi_j}\, \left (
h_{k-j}+[1+\frac{1}{k-1}] V_{k-1-j}\right )
\nonumber\end{eqnarray}
\vspace{-.4cm}
\beq
 =  L_{\chi_k} h+\frac{1}{k} V_{k-1}+
\sum^{k-1}_{j=1}\frac{j}{k}L_{\chi_j}\, \left (h_{k-j}+V_{k-1-j} \right )
\eeq
where the second term of the last equation 
is a consequence of   the recursive system for the coefficients $V_j$.
The term in the last bracket  is just $H_{k-1}$. In view of the commutator 
property
\beq
L_{\chi_k}\,h \equiv -L_{h}\,\chi_k
\eeq
we arrive at the desired recursion system (19)
\beq
L_h\,\chi_k + H_k=\frac{1}{k} V_{k-1}+
\sum^{k-1}_{j=1}\frac{j}{k}L_{\chi_j}\, H_{k-j}; \hspace{0.5cm} k=2,3,.. 
\eeq

\section*{Appendix B:\hspace{0.5cm} Integrable second order form of
the H\'{e}non-Heiles Hamiltonian}

We start from the transformed Hamiltonian (35), neglect the remainder
$V^{(3)}$, and write at first the resulting Hamiltonian $H^{(2)}_{trunc}$
in terms of action angle variables as defined in (37). We will abbreviate
 the   constant of motion $J_1'$ by $J$. When the action angle
variables $(J_2',\psi_2')$ are expressed by (28) in terms of cartesian
symplectic magnitudes $(p,q)$, we can write after some efforts
\beq 
H^{(2)}_{trunc}=J+\epsilon^2 [-\frac{5}{96}J^2+\frac{7}{48}J \, q^2-
\frac{7}{96}\,q^2 (p^2+q^2) ].
\eeq
This is one-degree of freedom Hamiltonian which can be brought into
integrable form in a standard way, see e.g. \cite{landau}.
The corresponding, exact canonical transformation from $(p,q)$
to action angle variable $(I,\Phi)$ is found as
\beqa
p&=&-2 \cos(\Phi)\sqrt{\frac{(J-I)I}{J-2\sin(\Phi)\sqrt{I(J-I)}}};
\nonumber\\
q&=&\frac{2I-J}{\sqrt{J-2\sin(\Phi)\sqrt{I(J-I)}}};
\hspace{0.3cm} J\geq I.
\eeqa
 With this we achieve the integrable form
\beq
H^{(2)}_{trunc}=J+\frac{\epsilon^2}{48} \left (14 I^2-14\,  I\,J+ J^2
\right )
\eeq  
in terms of  the action variables $J$ and $I$.


\begin{thebibliography}{99}

\bibitem{antonio} Giorgilli A. and Galgani L., Celest. Mech. {\bf 37} (1985),
95 

\bibitem{rasetti}  Rasetti M., {\it Modern methods in 
equilibrium statistical mechanics}, World Scientific (1986), Singapore

\bibitem{henon} H\'{e}non M. and Heiles C., Astron. J. {\bf 69} (1964), 73 

\bibitem{gustavson} Gustavson F.G., Astron. J.  {\bf 71} (1966), 670 

\bibitem{nekhoroshev} Nekhoroshev N.N., Usp. Mat. Nauk. {\bf 32}
(1977), 5 (Engl. transl. Russ.Math.Surv. {\bf 32 },1 (1177))

\bibitem{nekhoroshev2} Nekhoroshev N.N., in: Oleinik O.A. (Ed.),
 {\it Topics in Modern Mathematics, Petrovskii Seminar Nr.5 },
 Plenum Press (1979), New York 

\bibitem{deprit} Deprit A., Celest. Mech.  {\bf 1} (1969), 12 
 
\bibitem{galgani} Benettin G., Galgani L., and Giorgilli A.,
 Celest. Mech. {\bf 37} (1985), 1

\bibitem{rauh} Rauh A., Celest. Mech. {\bf 55} (1993), 161; Erratum,
ibidem p.415

\bibitem{niederman} Niederman L., {\it Stability over exponentially long
 times in the planetary problem} \linebreak
 preprint, Universit\'{e} Paris IX (1995)  

\bibitem{yoshida} Kinoshita H., Yoshida H., and Nakai H.,
 Celest. Mech. {\bf 50} (1991), 59

\bibitem{arnold} V. I. Arnold, {\it Geometrical Methods in the Theory
of Ordinary Differential Equations}, Springer-Verlag (1983), New York


\bibitem{rak} Rauh A., Andrade R.F.S., and Kougias F.Ch.,
{\it Symmetry in the normal form of the H\'{e}non-Heiles Hamiltonian}
in: Singular behavior and nonlinear dynamics, eds. St. Pnevmatikos, 
T. Bountis, Sp. Pnevmatikos, World Scientific (1989), Singapore 

\bibitem{holmes} Holmes P., Phys. Rep. {\bf 193} (1990), 137

\bibitem{gg} Giorgilli A., and Galgani L., Celest. Mech. {\bf 17} (1978),
267 

\bibitem{landau} Landau L.D. and Lifschitz E.M., {\it Mechanik}
(Akademie-Verlag (1981), Berlin  

\end{thebibliography}
\end{document}